\documentclass[pdflatex,sn-mathphys]{sn-jnl}
\usepackage{physics}

\jyear{2024}

\begin{document} 
\title[Optimal-order Trotter-Suzuki decomposition]{Optimal-order Trotter-Suzuki decomposition for quantum simulation on noisy quantum computers}

\author[1]{\fnm{A. A.} \sur{Avtandilyan}}\email{avtandilian.aa@phystech.edu}
\author*[2,1,3]{\fnm{W. V.} \sur{Pogosov}}\email{walter.pogosov@gmail.com}

\affil[1]{\orgname{Moscow Institute of Physics and Technology}, \orgaddress{\city{Dolgoprudny}, \postcode{141700}, \country{Russia}}}
\affil[2]{\orgname{Dukhov Research Institute of Automatics (VNIIA)}, \orgaddress{\city{Moscow}, \postcode{127030}, \country{Russia}}}
\affil[3]{\orgdiv{Institute for Theoretical and Applied Electrodynamics}, \orgname{Russian Academy of Sciences}, \orgaddress{\city{Moscow}, \postcode{125412}, \country{Russia}}}

\abstract{
The potential of employing higher orders of the Trotter-Suzuki decomposition of the evolution operator for more effective simulations of quantum systems on a noisy quantum computer is explored. By examining the transverse-field Ising model and the XY model, it is demonstrated that when the gate error is decreased by approximately an order of magnitude relative to typical modern values, higher-order Trotterization becomes advantageous. This form of Trotterization yields a global minimum of the overall simulation error, comprising both the mathematical error of Trotterization and the physical error arising from gate execution.
}


\maketitle

\section{Introduction}

When simulating quantum systems, classical computers encounter the exponential growth of the Hilbert space relative to system size. Manin \cite{manin1980computable} and Feynman \cite{Feynman} proposed that the inherent properties of quantum simulators or computers make them more natural tools for such investigations compared to classical computers. Presently, there is considerable interest in simulating spin systems, as they serve as a platform for exploring various physical phenomena, including quantum magnetism and chemical bonds in molecules \cite{ReviewQuanSim, QCh_energy, Antiferromagnet, HartreeFock, van_Diepen_2021, QuantChem}.

The primary approaches utilized for quantum simulation on digital quantum computers involve the variational principle and Trotterization of the evolution operator \cite{QS_state}. The variational method relies on assuming an ansatz for the ground state wave function, generated directly on the quantum chip, to approximately determine the energy of the ground state of the system \cite{Variation}. However, this approach may yield different results depending on the specific type of variational function employed \cite{Yuan_2019} and is susceptible to the issue of so-called barren plateaus \cite{VQABar}. Trotterization, on the other hand, involves decomposing the evolution operator into a series of unitary operations using the Trotter-Suzuki decomposition, enabling its mapping onto the quantum gates of a quantum computer. This method enables the simulation of system dynamics from a given initial state and the determination of ground state energy using a phase estimation algorithm \cite{kornell2023improvements}.
Recently it was shown that state-of-the-art IBM superconducting quantum computers can already simulate dynamics of spin systems using first-order Trotter-Suzuki decomposition \cite{IBM_Utility} which still remains a major challenge for classical computers\cite{protivIBM}.

The Trotter-Suzuki decomposition entails dividing the time interval of system evolution into a number of $n$ elementary time layers or Trotter steps \cite{Lloyd}. This decomposition becomes increasingly accurate as the number of layers approaches infinity, eliminating mathematical errors associated with discretization. However, an increase in $n$ leads to a rise in physical error stemming from quantum gate errors and other imperfections of quantum machines. Consequently, the total error reaches a minimum at an optimal number of Trotter layers. While this phenomenon has been demonstrated for first-order Trotter-Suzuki decomposition in previous studies \cite{Knee_2015,poulin2014trotter}, to our knowledge, the utility of higher-order Trotter-Suzuki decompositions \cite{HighOrder} that account for quantum gate errors remains unexplored. In this paper, we address this question and examine whether there exists a global minimum of the total error, comprising mathematical Trotterization error and physical gate execution error, with respect to the order $k$ of the Trotter-Suzuki decomposition. As illustrative examples, we investigate the dynamics of one-dimensional spin systems, focusing on two models: the Ising model in a transverse field and the XY model \cite{Ising_and_XY}. We argue that employing higher orders of the Trotter-Suzuki approximation is advantageous when the quantum gate error is sufficiently small, approximately an order of magnitude lower than in modern quantum processors.

\section{Trotter-Suzuki decomposition}

Let the Hamiltonian of the system of $N$ spins be given by
\begin{equation}
H= \sum_{i=1}^{K}H_i,\end{equation}
where each $H_i$ is a local operator, with $K$ being the number of these terms. Assuming $\hbar=1$, the dynamics of the system is described by the evolution operator
\begin{equation}\label{eq:u}
U(t)=\exp{-i tH} = \exp{-it \sum_{i=1}^{K} H_i}.
\end{equation}
This operator cannot be directly mapped into quantum gates and therefore requires decomposition. The first-order Trotter-Suzuki approximation is given by
\begin{equation}\label{eq:u1}
U_1(t) = \left( \prod_{i=1}^K e^{-\frac{it}{n}H_i} \right)^n.
\end{equation}
Effectively, it splits the evolution time into $n$ steps $U(t)=U(\frac{t}{n})^n$ and then decomposes the exponential of the sum into the product of exponentials. If $H_i$ are products of Pauli operators acting on different spins (qubits), which is often the case for many spin models, this product in Eq. (\ref{eq:u1}) can be directly mapped into the quantum circuit. In the third-order Trotter-Suzuki approximation, the evolution operator is given by
\begin{equation}\label{eq:u2}
U_3(t) = \left( \prod_{i=1}^K e^{-i\frac{at}{n}H_i} \prod_{i=K}^1 e^{-i\frac{\overline{a} t}{2n}H_i} \right)^n,
\end{equation}
where $a=1-\overline{a}=(3\pm \sqrt{3} i)/6$ . The general form of the decomposition is determined by the recursive relation \cite{SUzuki}
\begin{equation}\label{eq:u2k}
U_{k}(t)= U_{k-1}\left(p_k t\right) U_{k-1}\left(\left(1-p_k\right) t\right),
\end{equation}
with $p_k=(1+\exp (i \pi / k))^{-1}$. 

\section{Error estimation}

To determine the optimal Trotterization order, we evaluate both the mathematical error introduced by the decomposition and the physical error arising from imperfections in the quantum gates of the noisy quantum computer.

\subsection{Gate error}

For simplicity, we base our gate error estimation on the assumption that each exponential term $e^{-i\frac{t}{n}H_i}$ in the Trotter-Suzuki decomposition corresponds to a single gate in the quantum circuit, all with the same error probability $P_0$. While in reality, the mapping depends on the local Hamiltonian's type and the quantum computer's basis gates, resulting in different error rates for single-qubit and two-qubit gates, for the purpose of illustrating the advantage of higher decomposition orders, it suffices to assume uniform error across all operations.

In our model, we assume errors are independent, and the total fault rate is the sum of fault rates for each operation corresponding to the exponentials in Eqs. (\ref{eq:u1})-(\ref{eq:u2k}). Hence, for the first-order Trotter-Suzuki approximation, the fault rate is given by $r_{1}^{(g)}=KnP_0$, where $Kn$ represents the total number of gates in the quantum circuit. The probability of correct execution of the quantum circuit is $\exp(-r_{1}^{(g)})$, resulting in a probability of incorrect execution equal to $r_{1}^{(g)}$ in linear approximation. For the third-order Trotter-Suzuki decomposition, we have $r_{3}^{(g)}=2KnP_0$, and in the general case
\begin{equation}\label{eq:rg}
r_{k}^{(g)}=  2^{k-2} Kn P_0.
\end{equation}

Even though gate errors can accumulate in different forms, here we assume the worst case scenario, where all error accumulation is just addition of individual operations. This assumption is in a good agreement with recent simulations in Ref.\cite{IBM_Utility,TrotterError}.

\subsection{Mathematical error of Trotter-Suzuki decomposition}

Now, let us estimate the mathematical error arising from the Trotter-Suzuki decomposition of the evolution operator. In the leading order with respect to the deviation between the exact and Trotterized evolution operators, we can obtain the infidelity, which represents the probability of obtaining an incorrect result after the Trotterization, as follows:
\begin{equation}
1- \lvert \bra{\psi_0}U(t) ^{\dagger}U_k(t)\ket{\psi_0} \rvert^2 \simeq \bra{\psi_0}U(t) ^{\dagger}(U(t)-U_k(t))\ket{\psi_0} + \text{c.c.},
\end{equation}
where $\ket{\psi_0}$ is the initial state.

For small values of $t$, we can utilize a Taylor expansion to decompose the exponentials in Eq. (\ref{eq:u}) and Eqs. (\ref{eq:u1})-(\ref{eq:u2k}). For the first order, this results in the infidelity due to Trotterization $r_1^{(T)}$ as a sum of commutators of local Hamiltonians (see, e.g.,  Ref. \cite{Knee_2015}):
\begin{equation}\label{eq:r1}
r_1^{(T)} \sim  \frac{t^2}{2n} \langle\sum_{i \neq j} [H_i,H_j] \rangle,
\end{equation}
where $\langle \ldots \rangle$ refers  to the averaging over the state $\ket{\psi_0}$.
We estimate the infidelity $r_k^{(T)}$ corresponding to higher orders of decomposition as (see Theorem 6 and Corollary 7 in Ref. \cite{ErreorCommutator} for derivation of this equation):
\begin{equation}\label{eq:com}
r_k^{(T)} \sim \frac{t^{k+1}}{n^k} \langle \sum_{i_1 \neq ... \neq i_{k+1}}\left[H_{i_1}, \ldots[\ ] \ldots H_{i_{k+1}}\right] \rangle f(k),
\end{equation}
where $f(k)$ denotes some coefficient of order $1/k!$ that arises from the Taylor series. 

Next, we need to minimize the total probability of obtaining an incorrect result, which is a sum of physical error and mathematical error. We will now apply the derived formulas to two examples of one-dimensional spin systems: the transverse-field Ising model and the XY model.

\subsubsection{Transverse-field Ising model}

Let us start with the Hamiltonian of the one-dimensional transverse-field Ising model for $N$ spins:
\begin{equation}
H=-J\sum_{i=1}^{N-1}Z_iZ_{i+1} - \Gamma \sum_{i=1}^{N}X_i,
\end{equation}
where $Z$ and $X$ are Pauli operators, $J$ is the interaction constant between adjacent spins, and $\Gamma$ is the magnitude of the transverse field. In this case, the number of local Hamiltonians is $K=2N-1$. For the first order approximation in Eq. (\ref{eq:r1}), the only non-zero commutators will be $\left[Z_i Z_{i+1}, X_{i}\right]J \Gamma$ and $\left[Z_i Z_{i+1}, X_{i+1}\right]J \Gamma$. Since there are $N-1$ terms with $Z_iZ_{i+1}$, we have $2(N-1)$ identical contributions to the infidelity:
\begin{equation}
r_1^{(T)} \approx \frac{t^2}{2 n} 2 J \Gamma(N-1)=\frac{t^2}{n} J \Gamma(N-1).
\end{equation}

Taking into account Eq. (\ref{eq:rg}), the total error is:
\begin{equation}
r_1=r_1^{(T)}+r_1^{(g)} \sim P_0 K n+\frac{t^2}{n} N J \Gamma.
\end{equation}
For the one-dimensional Ising model $K=2N-1$ and for large $N$ we assume $ K \approx 2N$. The minimum is reached at $n^2 = J\Gamma t/(2P_0)$ and is equal to
\begin{equation}
r_1^{\min } \sim K t \sqrt{2P_0 J \Gamma}.
\end{equation}

For the second order decomposition ($k=2$), nonzero commutators are of the form $\left[\left[Z_i Z_{i+1}, X_{i}\right], X_{i} \right] J \Gamma^2$, $\left[\left[Z_i Z_{i+1}, X_{i+1}\right], X_{i+1} \right] J \Gamma^2$, and $\left[\left[Z_i Z_{i+1}, X_{i}\right], Z_i Z_{i \pm 1} \right] J^2 \Gamma$. So, the mathematical error in this case is:
\begin{equation}
r_2^{(T)} \sim \frac{t^3}{n^2} J \Gamma(4 J+4 \Gamma)(N-1) \sim 4 \frac{t^3}{n^2} J \Gamma(J+\Gamma) N.
\end{equation}
Together with the quantum gate error, we obtain the total error of the second order decomposition as:
\begin{equation}
r_2 \sim 2 \cdot P_0 K n+4 \frac{t^3}{n^2} N J \Gamma(J+\Gamma).
\end{equation}
The minimum is attained at:
\begin{equation}
r_2^{\min }= K t (2P_0)^{2 / 3}[J \Gamma(J+\Gamma)]^{1 / 3}.
\end{equation}

For the general case of the $k$-th order approximation, we estimate the mathematical error as:
\begin{equation}
r^T_k \sim \frac{t^{k+1}}{n^k}f(k) J \Gamma\left(2^k J^{k-1}+\ldots+2^k \Gamma^{k-1}\right) (N-1) \sim \frac{t^{k+1}}{n^k}f(k) J \Gamma(J+\Gamma)^{k-1} N A_1^k,
\end{equation}
where $A_1>1$ is a constant number and $f(k)$ is the same as in Eq.(\ref{eq:com}). Introducing $\Gamma =\alpha J$ we get the expression for the minimum total error as:
\begin{equation}
r_k^{\min } \sim t K P_0 J (\alpha+1) A_1 2^{k} \left(\frac{\alpha k f(k)}{(\alpha+1)^2 P_0 A_1}\right)^{\frac{1}{k+1}}.
\end{equation}

\begin{figure}[h]
    \centering
    \includegraphics[scale=0.23]{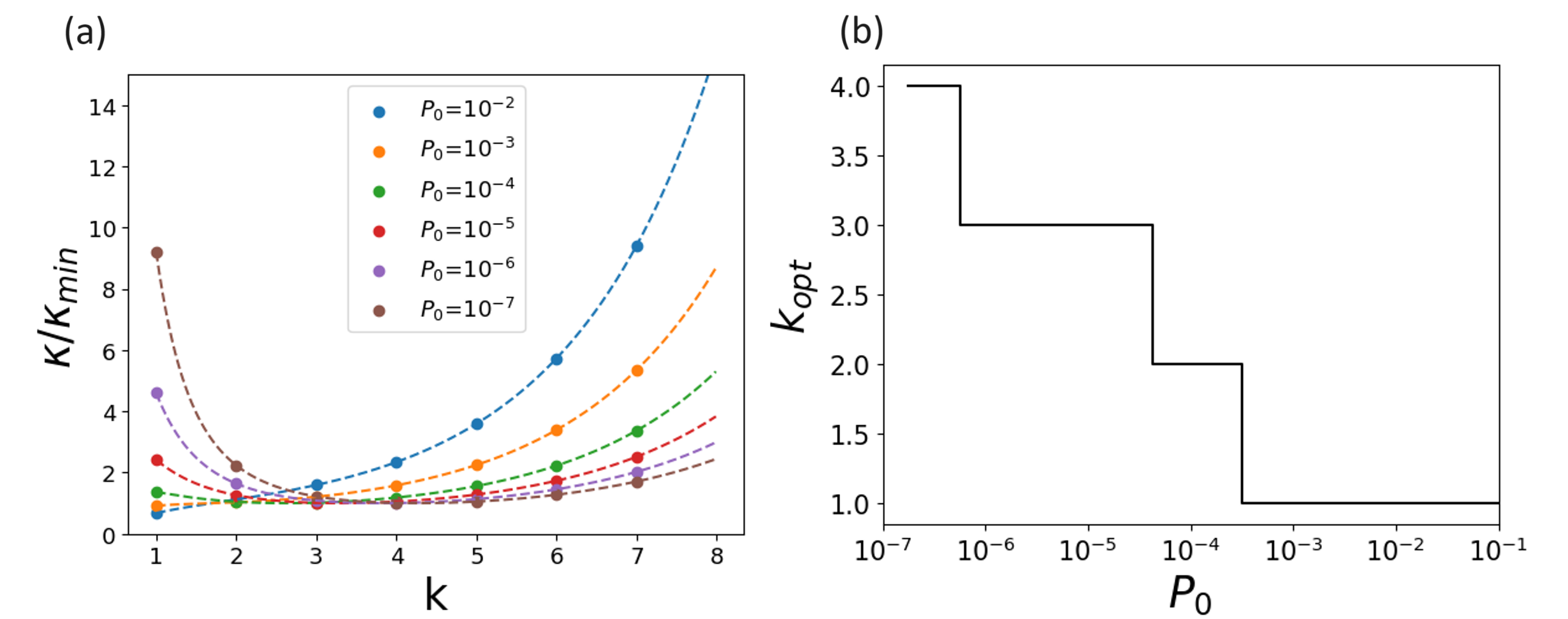}
    \caption{(a) Normalised $\kappa(k)$ dependence for different gate error values. Dashed lines are guides for eyes connecting points, which correspond to integer values of $k$. (b) Optimal decomposition order $k_{opt}$ for different gate errors $P_0$. The controlling parameters are $\alpha=1$, $A_1=2$.}
    \label{fig:Isingmin}
\end{figure}

The dependence of minimum total error on number $k$ can be expressed in a function $\kappa(k)=\left(B f(k) k \right)^{\frac{1}{k+1}}$ where we introduced $B=\alpha/(P_0 A_1 (\alpha+1)^2)$. We plot the function $\kappa(k)$ for different values of the gate error $P_0$ in Fig.\ref{fig:Isingmin}a) (specific values of $\alpha$ and $A$ do not qualitatively change our results). One can see that for high values of gate error (approximately $P_0 \gtrsim 10^{-4}$) function has no minimum and hence the optimal decomposition order will be $k_{opt}=1$. At lower values of gate error we see the formation of minima at higher orders. In Fig.\ref{fig:Isingmin}b) the dependence of optimal order of Trotter-Suzuki decomposition is plotted over gate error, showing that at
 $P_0 \sim 10^{-3}-10^{-4}$  it becomes favorable to use the second order decomposition.

\subsubsection{XY model}

\begin{figure}[h]
    \centering
    \includegraphics[scale=0.23]{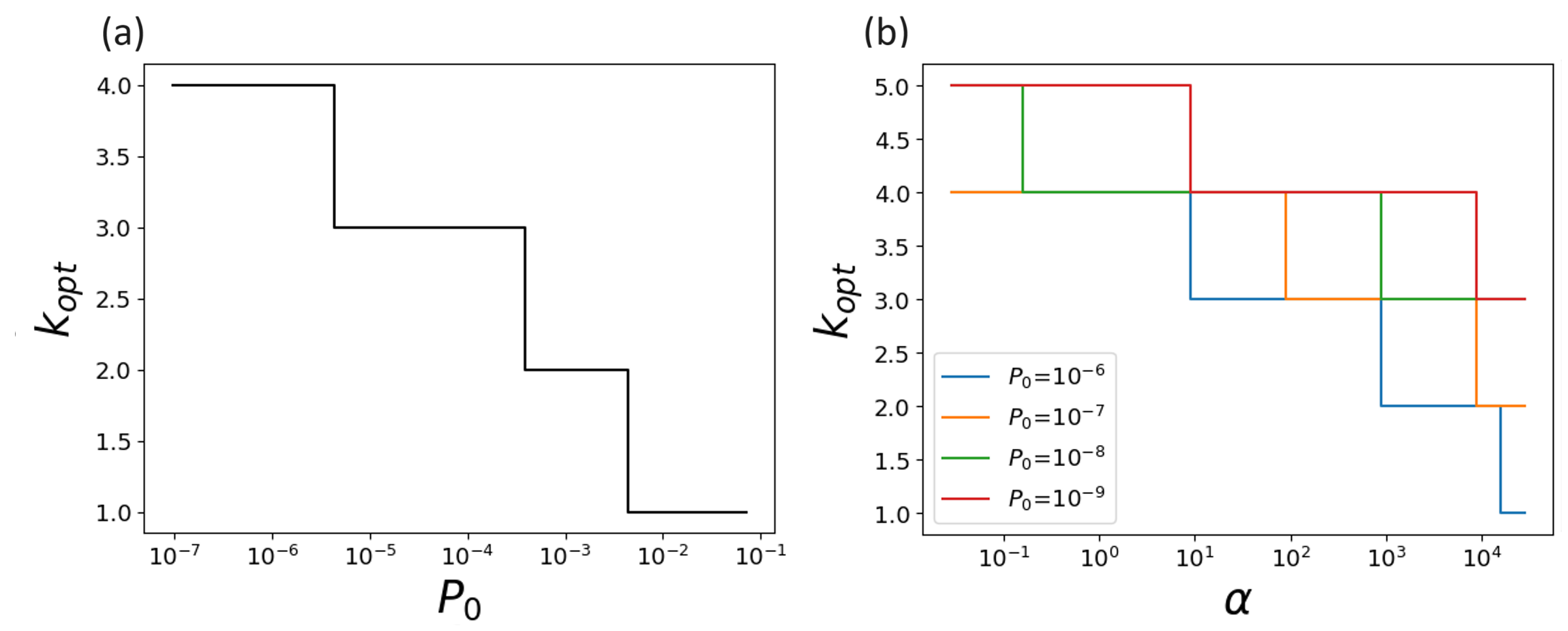}
    \caption{(a) Optimal order of Trotter-Suzuki decomposition $k_{opt}$ as a function of gate error. The values of controlling parameters are $\alpha=1$, $A_2=2$. (b) Optimal decomposition order as a function of parameter $\alpha$ for various values of gate error $P_0$.}
    \label{pl:XYmin}
\end{figure}

Let us now consider the one-dimensional XY model. The Hamiltonian is given by:
\begin{equation}
H=-J \sum_{i=1}^{N-1} X_i X_{i+1}-J \sum_{i=1}^{N-1} Y_i Y_{i+1}-\Gamma \sum_i^N X_i.
\end{equation}
Using Eq. (\ref{eq:com}), we can estimate the Trotterization error. For the first order of decomposition, the non-zero commutators are $ \left[ Y_i Y_{i+1}, X_{i} \right ] J \Gamma, \ \left[ Y_i Y_{i+1}, X_{i+1} \right ] J \Gamma, \ \left[Y_i Y_{i+1}, X_{i+1} X_{i+2}\right] J^2 \text { and }\left[Y_i Y_{i+1}, X_{i-1} X_i\right] J^2 $. This results in the mathematical error:
\begin{equation}
r_1^{(T)} = 2(N-1) J \Gamma+2(N-1) J^2.
\end{equation}
A detailed analysis of non-commuting operators for $k=2$ gives the Trotterization error as:
\begin{equation}
r_2^{(T)}= 4(N-1) J (\Gamma + 2J)^2.
\end{equation}
For arbitrary $k$, we estimate the Trotterization error for the XY model as:
\begin{equation}
r_k^{(T)} \sim \frac{t^{k+1}}{n^k} J(J+2 \Gamma)^k N A_2^k f(k),
\end{equation}
where $A_2$ is some number and $f(k)$ is $\sim 1/k$. Using Eq. (\ref{eq:rg}), we evaluate the optimal number of Trotter steps and the Trotter-Suzuki decomposition order $k_{opt}$. The results for $k_{opt}$ as a function of $P_0$ are presented in Fig. \ref{pl:XYmin} (a). In Fig. \ref{pl:XYmin} (b), we also show how $k_{opt}$ changes with the parameter $\alpha =\Gamma/J$.

\section{Discussion and Conclusion}

As depicted in Fig. \ref{fig:Isingmin} (a), the error function $\kappa(k)$ exhibits broad minima shifting towards larger values of the decomposition order with a decrease in gate error. This trend is further elucidated in Fig. \ref{fig:Isingmin} b), where the optimal decomposition order is plotted against the gate error. Importantly, in practice, we are concerned solely with integer values of the parameter $k$. Thus, we observe that there exists a critical gate error value $P^\prime_0 $, beyond which it becomes advantageous to utilize $k=2$. We estimate this critical value to be approximately $P^\prime_0 \approx 10^{-4}$. A similar pattern emerges for the optimal decomposition order in the XY model, where the critical gate error is estimated to be $P^\prime_0 \approx 10^{-3}$. Additionally, we demonstrate that the optimal decomposition values depend upon model parameters. As illustrated in Fig. \ref{pl:XYmin} (b), an increase in the ratio $\alpha=\Gamma/J$ makes it more favorable to employ a lower decomposition order to minimize the global error. As mentioned before, our estimates give upper bounds for gate and algorithmic errors \cite{TrotterBounds}. Moreover, the Trotterization error can be reduced using different orderings \cite{TrotOrdering} and error mitigating algorithms \cite{ErrorMitigating} for specific models. However, in our paper we would like to stress out the main point that overall systematic trend of reducing gate error $P_0$ shifts optimal order of decomposition to higher values.

Our findings underscore that an enhancement in qubit quality or the implementation of partial error correction, which reduces gate error by approximately one order of magnitude (compared to best modern values for two-qubit gates \cite{IonQubit}), renders higher orders of the Trotter-Suzuki decomposition advantageous for attaining more accurate quantum simulation results.

\section{Acknowledgments}

W. V. P. acknowledges support from the RSF grant No. 23-72-30004 (https://rscf.ru/project/23-72-30004/). 


\backmatter

\bibliography{refs}

\end{document}